\documentclass{iaus}
\usepackage{graphicx}

\title[Shells in the Magellanic System] 
{Shells in the Magellanic System}

\author[S. Stanimirovi\'{c}]   
{Sne\v{z}ana Stanimirovi\'{c}\footnote{Present address: 
Department of Astronomy, University of Wisconsin,
  475 North Charter Street, Madison, WI 53706, USA \break email: 
sstanimi.at.astro.wisc.edu}}

\affiliation{Radio Astronomy Lab, UC Berkeley, 601 Campbell Hall,
Berkeley, CA 94720, USA\\[\affilskip]}

\pubyear{2007}
\volume{xxx}  
\pagerange{119--126}
\date{?? and in revised form ??}
\jname{Proceedings of the IAU Symposium NO. 237}
\editors{Bruce G. Elmegreen  \& Jan Palous}

\def\kms{km~s$^{-1}$}
\begin{document}

\maketitle

\begin{abstract}
The Magellanic System harbors $>800$ expanding shells of neutral hydrogen,
providing a unique opportunity for statistical investigations.
Most of these shells are surprisingly young, 2--10 Myr old, 
and correlate poorly with young stellar populations. 
I summarize what we have learned about shell properties and 
particularly focus on the puzzling correlation between the shell radius 
and expansion velocity.
In the framework of the standard, adiabatic model for shell evolution
this tight correlation suggests a coherent burst of star formation across 
the whole Magellanic System. 
However, more than one mechanism for shell formation may be taking
place. 
\keywords{ISM: bubbles, ISM: structure, Magellanic Clouds}
\end{abstract}

\firstsection 

\section{Introduction}

Numerous studies over the past three decades have shown that 
shell-like structures dominate the interstellar medium (ISM) in many galaxies.
This sculpturing of the ISM, mainly assumed to be due to star-formation
activity, must have an imprint on many physical processes (e.g. the
transport of radiation, heating and cooling etc.).  
In the traditional scenario, shells are expected to be reservoirs of 
hot gas and are
the sole-source of the hot, intercloud medium. 
At the same time, these dynamic features are sites of local energy 
depositions in the ISM
which contribute significantly to the total energy budget.
Another important role shells have is in providing the connection between
galactic disks and halos. Shells grow in the disks, expand, and the largest
ones can reach sizes larger than the disk scale height. When this happens
shells open up and vent hot gas into the halo.

Despite significant observational efforts, 
the exact mechanism(s) for shell formation is still not fully
understood. 
Similarly, the late stages of shell evolution have not been explored. 
The most accepted model for shell formation, 
the ``Standard Model" (Weaver et al. 1977;
McCray \& Kafatos 1987), views shells as products of combined effects 
of stellar winds and supernovae. 
Numerous observational puzzles however motivated other suggestions.
For example, several other types of powering sources were suggested, 
the most exotic ones being pulsars and gamma-ray bursts. 
Several proposed mechanisms do not even require the existence of a central
energy source. These range from the collision of high-velocity
clouds (HVCs) with a galactic disk, through results of the 
general ISM turbulence, with or without gravitational and thermal
instabilities, the ram pressure stripping,
flaring of  radio lobes, to the complex new conceptual designs of the ISM in
the form of an elastic polymer interwoven with magnetic field lines (Cox
2005). This list is obviously long, and surprisingly keeps growing steadily.

In this paper we focus on shells discovered in the Magellanic System: the
Small Magellanic Cloud (SMC), the Large Magellanic Cloud (LMC), and the
Magellanic Bridge (MB). More than 800 shell-like structures were found and
cataloged in high-resolution neutral hydrogen (HI) observations of these three
environments. All observations were conducted with the
Australia Telescope Compact Array (ATCA) and were complemented with
the short-spacing data from surveys with the Parkes telescope.
These data sets sample a wide and continuous range of spatial scales, providing
unique opportunities for finding and studying large samples of expanding
shells. It is important to stress that
the three systems we investigate here probe very different 
interstellar environments.
The SMC is a dwarf irregular, gas-rich galaxy with a large line-of-sight
depth, the LMC is a dwarf disk galaxy with traces of spiral
structure and a higher star formation rate, while the MB is a column of gas between
the two galaxies that was formed as a result of tidal interactions. 
In Section 2 we summarize the most important properties of HI shells in the
Magellanic System, and address their implications in Section 3. In Section 4
we point out recent observational work in the domain of the late stages of 
shell evolution, and then summarize in Section 5.

\section{Summary of observational properties}
\label{sec:greenfun}

\subsection{Shells in the SMC}

\begin{table}\def~{\hphantom{0}}
 \begin{center}
  \caption{Summary of shell properties: Number of HI shells, the min/max range of
    shell radii, the min/max range of shell expansion velocities, the estimated
    mean dynamic age, and power-law slopes of the shell size and expansion
    velocity distribution functions.}
  \label{tab:kd}
  \begin{tabular}{lcccccc}\hline
    & N  & $R_{\rm s}$ &$V_{\rm exp}$
    & $\langle T_{\rm s} \rangle$ & $\alpha_{\rm r}$ &$\alpha_{\rm v}$   \\
    &    &(pc)   & (\kms) & (Myr) &  &  \\\hline
SMC  &509  &20--800  & 2--33  &5.7  &$-2.5\pm0.2$\rlap{$^a$}   &$-2.2\pm0.3$\rlap{$^a$} \\
 MB  & 163 & 10--200 & 2--20  &6.2  &$-3.6\pm0.4$   &$-2.6\pm0.6$    \\
LMC  & 124 (54)$^{b}$ & 50--620&6--36   & 4.8 &$-2.5\pm0.4$   &?   \\\hline
  \end{tabular}

$^{a}$ These slopes were derived using all SMC shells.\\
$^{b}$ The number in brackets refers to stalled shells.
 \end{center}
\end{table}

More than 500 expanding shells have been cataloged in the
SMC (Staveley-Smith et al. 1997; Stanimirovi\'{c}  et al. 1999; 
Hatzidimitriou et al. 2005). 
Table 1 lists the typical shell radius ($R_{\rm s}$) and the 
expansion velocity ($V_{\rm exp}$).
Curiously, as noted first by Staveley-Smith et al. (1997), all
SMC shells appear to have a very similar dynamic age, $T_{\rm s}\sim5$ Myr. 
This was interpreted as evidence for a single, coherent, and global 
burst of star formation in the SMC.
The volume occupied by all these shells is large, about 40\% of the whole SMC, 
implying a very bubbly, or a `Swiss Cheese'-like morphology.
Estimating the fraction of HI mass  occupied by shells is more 
difficult as we need an
estimate of the local ambient density. If we
assume that the local ambient surface density for {\it all shells} is 0.01 M$_\odot$
pc$^{-2}$, we arrive at a total mass fraction that is about 20\%. 

Recently, Hatzidimitriou et al. (2005) searched for the remnant stellar 
population
associated with the HI shells by using all available catalogues of the young
stellar populations.
About $450$ shells were found to correlate with one or more stellar 
objects, while
59 shells do not correlate with any of the known stellar objects. 
We will refer further to these two classes of shells as 
``non-empty" and ``empty", respectively. 
The surprising result is that properties of ``non-empty" and ``empty" 
are almost indistinguishable.
There are no morphological differences between the two groups. 
``Empty" shells appear smaller and with a lower expansion velocity than
``non-empty'' ones, however this is primarily a selection effect. 
Both types also show an almost linear correlation between the shell radius 
and expansion velocity. 
Spatially, ``empty"  shells are found primarily in remote places on the
outskirts of the HI distribution. Of course, finding similar shells in the
central parts would be impossible. Several ``empty" shells with high
luminosity appear loosely clustered and possibly connected; these
objects may belong to an old chimney.  
The rest of the empty shells, however,
do not have any special location or association.

Hatzidimitriou et al. (2005)
also derived the shell size and expansion velocity distribution functions,
$N(R_{\rm s})$ and $N(V_{\rm exp})$. 
For both types of shells $N(R_{\rm s})$ and $N(V_{\rm exp})$ can be fitted 
with a power-law function, $N(R_{\rm s}) \propto R_{\rm s}^{{\alpha}_{\rm
    r}}$ and $N(V_{\rm exp}) \propto V_{\rm exp}^{{\alpha}_{\rm v}}$.
The slopes $\alpha_{\rm r}$ and $\alpha_{\rm v}$ agree within their uncertainties for 
the two types of shells.
In the framework of the standard model, based on whether all shells were 
formed in a  single burst or in a continuous manner, and by assuming 
either a single input mechanical luminosity function (MLF) for all shells or a 
power-law function, we can predict $\alpha_{\rm r}$ and
$\alpha_{\rm v}$ and then compare these values with what we 
get from observations.  This was first shown by Oey \& Clark (1997). 
The positive slope of the $R_{\rm s}-V_{\rm exp}$ correlation, 
and the fact that $\alpha_{\rm r} \approx \alpha_{\rm v}$
($\approx -2.2\pm0.2$ for ``non-empty shells),
point to the case of a single burst of shell formation and a 
power-law input MLF. 
It is puzzling though that the same arguments apply to ``empty" shells as
well.
If the ``empty" shells are $>10$ Myr old, and this could be the reason why we do not find
their corresponding stellar population, then for them to fit on the same
$R_{\rm s}-V_{\rm exp}$ relation would require a significant and 
concerted re-acceleration.

\subsection{Shells in the Magellanic Bridge}

Muller et al. (2003) cataloged 163 shells in the MB, applying criteria
somewhat tighter than in the case of the SMC shells (for example, they do not
include incomplete large shells in their catalog; this results in shell
sizes being biased towards smaller shells). 
Shell sizes and expansion velocities are given in Table 1.
The estimated mean dynamic age is $T_{\rm s}=6.2$ Myr, with a
standard dispersion of 3.4 Myr. Muller et al. (2003) cross-correlated 
their shell catalog with the catalog of OB associations by 
Bica \& Schmitt (1995) and found that about 60\% of shells do 
not have corresponding
 OB associations. Also, while the mean dynamic shell age is about 6 Myr, the
mean age of OB associations is several times larger, 10-25 Myr. 
Although the MB is a tidal remnant of the interactions between the 
LMC and the SMC,
shells in the MB are primarily spherical and without obvious signs of 
distortions or tidal stretching.

\subsection{Shells in the LMC}

There are 101 giant ($R_{\rm s}<360$ pc) and 23 super-giant 
($R_{\rm s}>360$ pc) shells in the LMC (Kim et al. 1999). 
Shell radii are in the range  50--620 pc, while shell
expansion velocities are in the range 6--36 \kms. Interestingly, 
while the expansion velocity is systematically higher in the LMC 
than in the SMC, 
about one half of all LMC shells appear to have stalled, with $V_{\rm exp}=0$. 
The mean dynamic age is 4.9 Myr, which is again younger than the age 
of corresponding OB associations ($>10$ Myr). 
Giant shells in the LMC also follow the almost linear $R_{\rm s}-V_{\rm exp}$ 
relation, while the super-giant shells deviate from this trend. 
Kim et al. (1999) also found a poor spatial correlation
between shells and OB associations.  
The shell size distribution is
$N(R_{\rm s}) \propto R_{\rm s}^{-2.5}$, and is similar to 
that for the SMC shells.

\section{Putting it all together}

\begin{figure}
\includegraphics[height=3.5in,width=5in,angle=0]{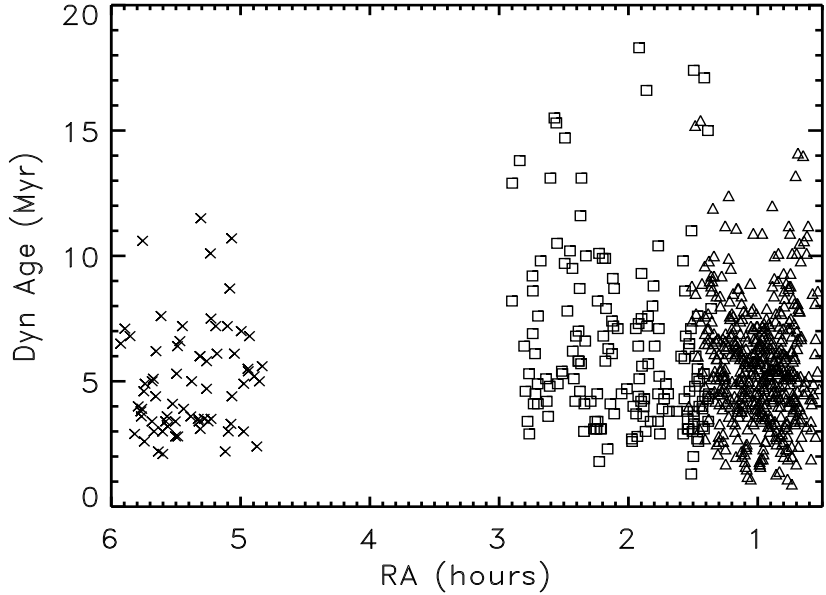}
  \caption{The estimated dynamic age for shells in the SMC (triangles), the MB
  (squares), and the LMC (crosses).}\label{f:T}
\end{figure}

\begin{figure}
\includegraphics[height=3.5in,width=5in,angle=0]{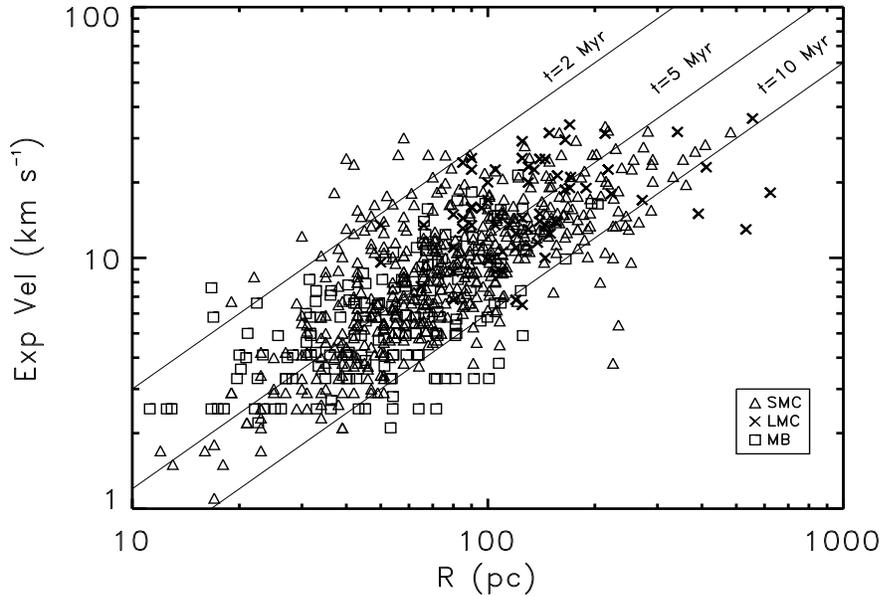}
  \caption{Size and expansion velocity for shells in the SMC (triangles), the
    LMC (crosses), and the MB (squares). The solid lines represent 
    dynamic ages of 2, 5, and 10 Myrs, respectively, in the framework of the
    standard model.}\label{f:R-V}
\end{figure}

To summarize, the properties of all shells in the SMC, the MB,
and the LMC show striking similarities: the dynamic age, tight
$R_{\rm s}-V_{\rm exp}$ relation, statistical properties, 
poor correlation with stellar populations,
and dynamic ages being younger than those of OB associations. And yet, the
three environments in which these shells formed and evolved are 
drastically different! 
To emphasize this, we plot the dynamic shell age as a function of 
Right Ascension 
in Figure~\ref{f:T} and include all ($>800$) shells found in the 
Magellanic System.
This distribution appears uniform and there are no obvious discontinuities 
between the SMC and the MB.
It is also apparent that the SMC and the MB have a few shells older 
than the majority of the LMC shells.
Figures~\ref{f:T} and \ref{f:R-V} also show that, on average, 
the shell size increases from the SMC, through the MB, to the LMC. 
This is primarily a selection effect, however. 
For example, Muller et al. (2003) discuss their lack of large shells in the MB
and explain that this is mainly due to their more stringent criteria when
identifying shells.

In Figure~\ref{f:R-V} we plot all shells on the $R_{\rm s}-V_{\rm exp}$
diagram. It is very curious how all
shells appear to follow the same, tight relation, being nested between 
the solid lines that mark $T=2$ Myr and $T=10$ Myr, when viewed 
within the standard model. In the case of our Galaxy, 
Ehlerov{\'a} \& Palou{\v s} (2005) did not find
a correlation between the shell radius and expansion velocity, although even
there it looks like larger shells have a larger $V_{\rm exp}$. 
In the framework of the standard model, the tight age
spread could be interpreted as a result of a recent star formation burst
across the whole Magellanic System about 5 Myr ago. However,
this is difficult to reconcile with the star formation history of the Clouds.
In the case of the SMC, Harris \& Zaritsky (2004) estimated ages of
$>5\times10^{6}$ stars and found 
the closest peak in star formation 60 Myr ago.
It is also interesting to note that on the older-age side, the cut-off in 
the shell distribution (Figure~\ref{f:R-V}) is sharper than on 
the younger-age side. 
This may be suggestive of an enhanced shell destruction/fragmentation after
$\sim10$ Myr. However, this is significantly shorter than the typical predicted 
shell lifetime of 30--50 Myr (Dove 2000; W\"{u}nsch \& Palou\v{s} 2001).

The $R_{\rm s}-V_{\rm exp}$ diagram obviously suffers from some selection
effects and is biased towards younger objects.  Large shells 
are hard to distinguish
observationally as they are often fragmented and also could be seen as a
superposition of several smaller shells.
In the case of the SMC, Zaritsky \& Harris (2004) 
showed that 10 to 70\% of all stars could have 
formed through tidal triggering.  
Tidal triggering would obviously affect shell formation, and probably
even shell evolution. The degree with which these processes 
are constrained and coordinated across the Magellanic System still 
needs to be explored. 
Alternatively, there may be one or more additional
shell-formation processes taking place. Obviously, understanding the role of 
environmental effects (e.g. tidal flows, the turbulent ISM, interactions
between shells, magnetic field etc.) is crucial for further advancement.

\section{What happens to shells at a very late stage of their evolution?}

We would now like to draw attention to numerous HI clouds found recently in the
interface region between the disk and the halo of our Galaxy 
(Lockman 2002; Stil et
al. 2005; Stanimirovi\'{c} et al. 2006).  These clouds are small, 5--20 pc in size,
cold ($<400$ to 1000 K), and often kinematically follow the Galactic disk but
at a velocity that is offset by 10--20 \kms~from that of the bulk HI emission.
The clouds do not appear to prefer particular regions in the Galaxy, their
distribution is most likely radially extended. The origin of these 
clouds is not clear. There are several possibilities including 
Galactic fountains and the
accretion of extragalactic gas. An alternative possibility, however, is 
that these clouds are fragments of expanding shells. 
There are several pieces of evidence that
point in this direction. For example, a Galactic chimney, GSH242-03+37 
shows small, discrete HI clouds that appear associated 
with shell caps (McClure-Griffiths et al. 2006). 
Lockman, Pidopryhora, \& Shields (2006) found a large plume-like 
structure which has 
numerous clouds. Stanimirovi\'{c} et al . (2006) 
found that clouds are embedded in large filamentary structures and
morphologically resemble cold clouds that form in simulations
of dynamically triggered instabilities (e.g. Audit \& Hennebelle 2005).
One way of testing the hypothesis that these newly-discovered clouds
are shell fragments is to derive the cloud mass spectrum and compare it with
theoretical predictions from shell fragmentation.
For example, W\"{u}nsch \& Palou\v{s} (2001) predict that the mass spectrum for shell
fragments is a power-law with $dN/dM \propto M^{-1.4}$.
The mass spectrum of observed clouds will be easily derived in the near
future as several large HI surveys are underway with the Arecibo, Parkes and
Green Bank telescopes that are particularly suited for finding these clouds.

\section{Summary and Open Questions}
There are $\sim800$ shells in total in the SMC, the MB and the LMC. 
At least 1/10 of SMC shells are devoid of  stellar counterparts, 
but surprisingly have properties similar to those of ``non-empty'' shells. 
Large similarities in shell properties across the Magellanic System,
and especially the tight correlation between the shell radius and expansion
velocity, are puzzling and may be highlighting the importance of 
tidal interactions for both shell formation and evolution. 
Alternative processes for shell formation and external effects
may be also playing important roles. The late stages of shell 
evolution are finally being addressed observationally, and 
detailed theoretical attention is highly desirable in this area.   
And finally, there is the Magellanic Stream, a starless tidal tail 
that provides the perfect opportunity to quantify the
importance of shell formation processes without a powering
source. This is an obvious future project!

\begin{acknowledgments}
I would like to thank the conference organizers for inviting me to participate
in this highly stimulating and enjoyable conference. 
I would also like to thank Carl Heiles, one of
the pioneers of shell exploration, for numerous inspiring
discussions, and Jacco van Loon for many insightful comments. 
This work was partially supported by NSF grants AST 04-06987 and 00-97417.
\end{acknowledgments}

\end{document}